\documentstyle[prl,aps,epsf]{revtex}
\advance\textheight by 0.14in
\advance\topmargin by -0.1in
\begin{document}
\draft

\twocolumn[\hsize\textwidth\columnwidth\hsize\csname@twocolumnfalse\endcsname

\title{Deconfinement and Dissipation in Quantum Hall ``Josephson'' Tunneling} 
 
\author{H.A.\ Fertig and Joseph P. Straley}

\address{Department of Physics and Astronomy,
 University of Kentucky, Lexington, Kentucky 40506-0055}

\date{\today}

\maketitle

\begin{abstract} 
The zero-bias tunneling resonance in quantum Hall bilayer systems
is investigated via numerical simulations of the classical two dimensional
$XY$ model with a symmetry-breaking field.  Disorder is included in
the model, and is shown to nucleate
strings of overturned spins proliferated through the system, with unpaired vortices
and antivortices at their endpoints.
This {\it string glass} state supports low energy
excitations which lead to anomalously large dissipation in tunneling,
as observed in experiment.  The effect of an in-plane magnetic field
is discussed.
\end{abstract}

]
{\it Introduction.} Bilayer quantum Hall systems have attracted
much attention because, for small layer separations and vanishing
interlayer tunneling, they support states with spontaneous 
interlayer coherence.  This broken symmetry leads to a
Goldstone mode\cite{fertig1} which may be
described as a spin wave in an
$XY$ ferromagnet\cite{review}.  If one introduces interlayer
tunneling, the system may alternatively 
be described as a large area
Josephson junction.  In the $XY$ language, the
tunneling appears as a magnetic field tending to
align the spins along the $\hat{x}$ direction-- i.e. an in-plane field.
While this symmetry-breaking field makes the physics of this
system completely different than the heavily studied model
without an in-plane field, vortex deconfinement does occur,
albeit not via the usual Kosterlitz-Thouless
mechanism\cite{fertig2,fertig3}.
The Josephson junction analogy
suggests these systems might support
a Josephson effect\cite{wen}.
Recently, a sharp conductance resonance
near zero interlayer voltage bias was detected\cite{spielman1,spielman2}
which is highly reminiscent of the DC Josephson effect. 
The width of this resonance, however, is surprisingly large, and perturbative
treatments\cite{balents,stern,fogler} of the system have so far
not been able to account for the dissipation.   As we demonstrate below,
the resolution of this mystery is likely related to the unbinding 
of vortices.

In this article, we report on the results of
extensive simulations of this
system,  both with and without disorder
(introduced via a random vector field described below), focusing
on the effects of vortex deconfinement.
Our results have important implications for this model
and its many realizations.  Most significantly,
when the 
disorder is sufficiently strong, a new type of low-temperature
state emerges in which strings of overturned spins
proliferate, with unpaired vortices at their endpoints.
Low-energy excitations appear that are localized at
the vortices, and the strings themselves behave in 
many respects like a pinned manifold.   These observations
are consistent with glassy behavior, so we call this
a {\it string glass} state.  
The string glass is likely
a ``frozen-in'' version of the deconfined state
found recently in studies of this system in the clean limit\cite{fertig2,fertig3}.
While we focus below on the implications
of this state for the bilayer quantum Hall system,
analogous states should also exist in planar Josephson
junctions, real planar magnets, and two-dimensional
crystals in an appropriate disorder environment,
with consequences analogous to those we find
for the quantum Hall system.

For tunneling in the bilayer quantum Hall system, our
results have a number of experimental implications.
Firstly, the localized excitations near
the vortices are directly excited by a tunneling current,
leading to an anomolously large dissipation at low
current.  Such dissipation is known to exist in 
real systems\cite{spielman1}, but its large magnitude so far has defied
explanation\cite{balents,stern,fogler,joglekar}, particularly its
tendency to have a weak temperature dependence relative to
what one might find if the dissipation is caused by quasiparticles.
By contrast, we find that dissipation associated with 
the vortices in the string glass state
intrinsically has a weak temperature dependence, 
as illustrated in Fig. 1.  
Above a critical
current scale $i_c$, we find that the strings depin 
and the resistance, as well as the voltage noise,
is greatly enhanced.  An observation of a peak in the noise
near  $i_c$ would give direct experimental evidence of the
existence of the string glass state posited here.
Finally, we show that
an in-plane magnetic field $B_{\parallel}$ 
tends to align the strings
along the field direction, and for large
enough  $B_{\parallel}$  the density of strings
increases.  At low temperature, the strings are pinned
by disorder, so the zero-bias resonance remains
in place.  Again, this has been observed experimentally\cite{spielman2},
but to our knowledge this is the first qualitative explanation
of why this occurs.

{\it Simulation Model. } To conduct our simulations
we place our system on an $L \times L$  square lattice, whose lattice constant
$a_0$ should be understood as the underlying microscopic length
scale of the system, the magnetic length $\ell_0=\sqrt{\hbar c/eB}$,
with $B$ the perpendicular magnetic field.
We impose periodic boundary conditions on this lattice.
Our Hamiltonian is $H=H_{XY} + H_D$, with
\begin{equation}
H_{XY}= -K \sum_{<{\bf r}{\bf r}^{\prime}>}
 \cos[\theta ({\bf r})-\theta ({\bf r}^{\prime})] - h\sum_i \cos{\theta ({\bf r})},
\label{XYmodel}
\end{equation}
\begin{equation}
H_D= \sum_{<{\bf r}{\bf r}^{\prime}>}{A_{{\bf r}{\bf r}^{\prime}}}
 \sin[\theta ({\bf r})-\theta ({\bf r}^{\prime})].
\label{disorder}
\end{equation}
In Eq. \ref{XYmodel}, $\theta ({\bf r})$ represents the angle
of a planar spin located
at lattice site ${\bf r}$, $\sum_{<{\bf r}{\bf r}^{\prime}>}$ denotes a sum over nearest 
neighbors, and $h$ represents the effect of a magnetic field
tending to align spins
along $\theta ({\bf r})=0$, whose underlying source is the interlayer
tunneling matrix element of the electrons.  In $H_D$,
$A_{{\bf r}{\bf r}^{\prime}}$ is a random vector field residing on the bonds,
which may be understood as follows.  At long wavelengths,
the state of the bilayer quantum Hall system may be specified
by a {\it three} dimensional pseudospin field ${\bf S}({\bf r})$.  $S_z$ 
here represents the difference in electron density between the two
layers, and $S_x,~S_y$ are the components of the planar spin field.
$S_z$  
ideally vanishes when the electric potential of the two
layers is perfectly balanced; in practice, there are
impurities which tend to push the electrons into one layer
or the other.  These same impurities will tend to locally lower or raise the
total electron density, which can be accomplished
by introducing a non-vanishing topological density
$q({\bf r})= \epsilon_{\mu\nu} {\bf S} \cdot [\partial_{\mu} {\bf S}
\times  \partial_{\nu} {\bf S}]/8\pi$,
which in the quantum Hall context is proportional to the
{\it real} charge density\cite{lee}.  The topological density
contains a term of the form $\vec{A} \cdot \vec{\nabla} \theta$
\cite{stern}, with $\vec{A}$ proportional to the induced $S_z({\bf r})$.
It is this contribution to the disorder we model, taking $\partial_{\mu} \theta
\rightarrow  \sin[\theta ({\bf r})-\theta ({\bf r}^{\prime})]$, with
 $<{\bf r}{\bf r}^{\prime}>$ the nearest neighbor
bond along the $\mu$ direction, to account for the finite grid spacing
in the simulation\cite{com}.  
We take $A_{\mu}({\bf r})=\sum_{{\bf r}^{\prime}}
e^{-|{\bf r}-{\bf r}^{\prime}|^2/\xi_p^2}v_{\mu}({\bf r}^{\prime})$,
with $v_{\mu}$ drawn from a uniform distribution satisying  
$-\Delta < v_{\mu} < \Delta$\cite{com1.5}.
Because disorder due to charged
impurities in most quantum Hall systems varies slowly in space, the
results we present are for relatively large values of $\xi_p$. 
Our qualitative results however also hold for smaller values of $\xi_p$. 

To simulate dynamics in this model, we adopt a Langevin equation of motion,
\begin{equation}
\Gamma {{d^2 \theta({\bf r}) } \over {dt^2}} = 
{{\delta H} \over {\delta \theta({\bf r}) } }
+ \zeta({\bf r}) - \gamma {{d \theta({\bf r}) } \over {dt}} + i_d.
\label{eqmot}
\end{equation}
$\Gamma$ above represents an effective moment of inertia for an $XY$ spin.
Since $\theta$ is conjugate to $S_z$ in the full quantum theory, $\Gamma$
originates from
the energy cost for creating charge imbalance across the layers, and is
proportional to the capacitance per unit area of the bilayer.  The $\zeta$
and $\gamma$
terms are respectively a random torque and a uniform damping, introduced
to model coupling to a heat bath.  To satisfy the fluctuation-dissipation theorem, the
random torques are drawn from a distribution satisfying
$<\zeta({\bf r},t)  \zeta({\bf r}',t')> = 2\gamma T \delta_{{\bf r},{\bf r}'} \delta (t-t')$
with $T$ the temperature of the system.
Finally,  the last term
is a current which may be imposed to drive the system\cite{barone}. 

{\it Results.} We begin by studying the resistance of the system near zero
bias.  This can be conveniently accomplished by examining the dynamics
of the $XY$ spins via the extended variables 
$u({\bf r},t) = \int dt {{\partial \theta({\bf r})} \over {\partial t}}$.
The variables $u$ are identical to the original angles except
they are unbounded.  Their dynamics are analogous to those of
particles in a washboard potential\cite{halperin}, and in the absence
of a drive current ($i_d=0$) they diffuse.   It is easily confirmed
that $<u^2({\bf r},t)> \propto t$ for large enough time $t$, where
$<\cdot >$ denotes an average over sites and disorder configurations.
We thus define a diffusion constant $D = {\rm lim}_{t \rightarrow \infty}
d<u^2({\bf r},t)>/dt$.   The ``conductivity'' of the $u$'s is then
given by the Einstein relation $\sigma = D/T$, which quantifies their
linear response to the drive current.  Since uniform
average motion of the $u$'s generates a (Josephson) voltage, this means
$\sigma$ is proportional to the tunneling resistance of the system, $R$.

\begin{figure}
 \vbox to 5.0cm {\vss\hbox to 10cm
 {\hss\
   {\includegraphics{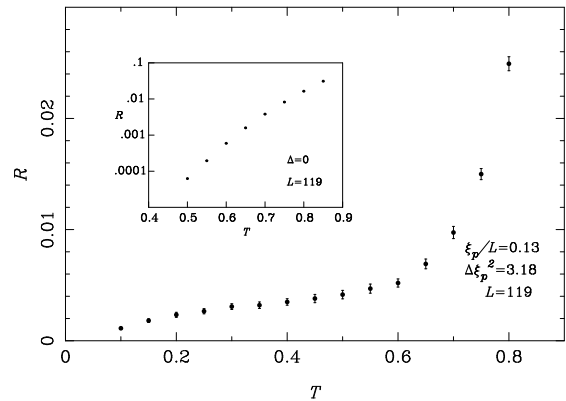}
   }
  \hss}
 }
\vspace{10mm}
\caption{Tunneling resistance averaged over 16 disorder realizations and
$10^6$  Langevin sweeps for
states in string glass state.  Length, time and energy units 
chosen so that  $\Gamma=K=a_0=1$;  $h\xi_p^2=.25$, $\gamma=0.143$ in these units.
Inset: Corresponding resistance for no disorder.}
\label{fig1}
\end{figure}

Fig. 1 illustrates some typical results for a measurement of this quantity in a series
of runs.  In these simulations, the system was annealed from above the Kosterlitz-Thouless
temperature ($T_{KT}=0.892K$), with temperature steps $\Delta T = 0.05$.
In the inset we show the result for a clean system, for which
the tunneling resistance drops exponentially at low temperatures.
At higher disorder levels (main
figure)  the low temperature resistance is increased
by many orders of magnitude, demonstrating the crucial role played by
disorder.

To see why this is so, one may examine the detailed behavior
of the $XY$ spins.  Fig. 2 illustrates a snapshot of a system configuration
at low temperature.  We show only spins for which $\cos \theta({\bf r}) < 0$;
i.e., spins whose $x$ component is opposed to the direction specified by
the symmetry breaking field $h$.  Qualitatively, one may see these form string-like
structures.  We find that narrow necks and endpoints of these strings
are often associated with isolated vortices \cite{fertig2,com5}.  Also
shown as circles are the 100 most active sites,
those with the greatest change in $u({\bf r})$
over the course of the simulation.   These are clearly located in the vicinity of the 
isolated vortices, demonstrating that the barrier for overturning spins in
such locations is greatly reduced compared to other locations in the system.
This is the underlying cause of the enhanced dissipation
in these systems.  At low temperatures and in the absence of a
drive current $i_d$, we find that the strings (vortices)
are pinned.  Physically we expect this because these objects
carry charge density (net charge) and so may be held in place by disorder.
Because of the quenched
nature of the disordered state, we call this a {\it string glass}.  
Low-energy, localized excitations are a common
feature of glasses\cite{ziman}, and their presence here
supports our characterization of this as a glassy state.\cite{com4}

\begin{figure}
 \vbox to 5.0cm {\vss\hbox to 10cm
 {\hss\
   {\includegraphics{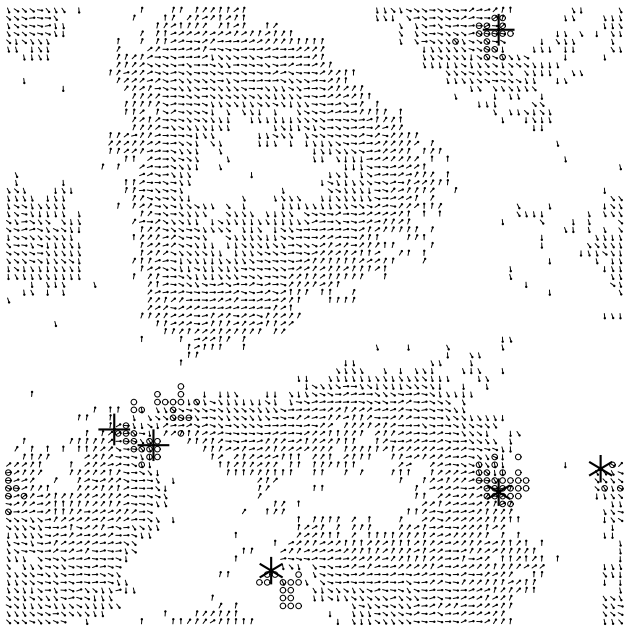}
   }
  \hss}
 }
\vspace{40mm}
\caption{Snapshot from run with $\Gamma=K=a_0=1$, 
$L=79$, $h\xi_p^2=.25$, $\gamma=0.143$, $\xi_p/L=0.13$,
$\Delta\xi_p^2=3.18$.  Only spins with $\cos \theta < 0$ shown.
Circles indicate most active spins, which are localized near vortices $(+)$ and
antivortices $(*)$}
\label{fig2}
\end{figure}

In the string glass state, a drive current acts as an effective force on the 
strings.  In a typical run, at very low drive the average ``displacement''
$\int d^2 r~u({\bf r},t)/L^2$ for a single
disorder realization has a small slope.  
As $i_d$ is increased, one finds occasional short intervals for which this slope is
increased, occuring when string sections become briefly depinned.
With further increasing $i_d$
these events become more frequent, until 
a critical drive current $i_c$ is reached at which
the strings essentially move freely. 
In this situation,
spins that were previously nearly static now
rotate much more frequently as strings pass through them (inducing
a Josephson voltage),
and the dissipation in the system is greatly enhanced \cite{url}.  This is apparent
in Fig. 3, which illustrates the current-voltage characteristic for a
single disorder configuration, with each data point averaged over eight
different initial seeds for the random force. For each of these seeds we have
averaged over $2 \times 10^6$ Langevin sweeps, with the first 25\% of
these thrown away for equilibration.  The onset of ``normal'' dissipation
in this system is apparently a depinning phenomenon.   Because of the
stick-slip motion in the vicinity of $i_c$ described above, there is considerable
broadband noise in the dissipation for $i_d \sim i_c$.  This is typical of a depinning
phenomenon, and should generate a peak in the noise spectrum of the power
dissipated.  {\it An experimental observation of such a noise peak would
confirm the string glass nature of the state, since it reflects the depinning
of the strings.}

An in-plane component of the magnetic field leads to further interesting
physics.  For an appropriate choice of gauge, 
the in-plane field introduces a phase
in the tunneling matrix element which causes the $XY$ spins to
tumble spatially with a wavevector 
${\bf Q}=(de/\hbar c){\bf \hat{z}} \times {\bf B}_{\parallel}$,
with $d$ the separation between layers and ${\bf B}_{\parallel}$
the in-plane component of the field.  This can be incorporated into
$H_{XY}$ by making the substitution
$ \cos[\theta ({\bf r})-\theta ({\bf r}^{\prime})]
 \rightarrow  \cos[\theta ({\bf r})-\theta ({\bf r}^{\prime})-Q_{{\bf r}{\bf r}^{\prime}}]$
in Eq. \ref{XYmodel}, where $Q_{{\bf r}{\bf r}^{\prime}}=Qa_0$ 
resides on nearest neighbor
bonds along the direction of ${\bf Q}$.  Note that because of the
periodic boundary conditions, we are constrained to choose values
of $Q$ such that $n_Q \equiv LQ/2\pi$ is integral.  
In the string glass state, for small
values of $n_Q$, the main qualitative effect we observe 
in configurations produced by simulated annealing is a tendency
for the strings to align along {\bf Q}. 
For larger  $n_Q$, we find the strings become narrower 
and the configurations appear more ordered, resembling a disordered
soliton lattice\cite{cote} -- with the strings being essentially
identical to solitons.

This qualitative transition is reflected in the critical current $i_c$,
as illustrated in Fig. 3.  One can see $i_c$ is little
changed from its $n_Q=0$ value for $n_Q=1,2$, whereas
for $n_Q=3$ and above, $i_c$ becomes noticeably suppressed.
The existence of a parallel field scale above which the depinning
current decreases can be understood in terms of collective
pinning theory\cite{larkin}.  For large values of
$Q$, the soliton lattice may be thought of as an elastic
medium, which is deformed and pinned by the disorder.
This deformation involves a length scale $L_c$ for a typical
domain size, determined by balancing the pinning and elastic
distortion energies\cite{larkin}.  
The length scale is meaningful provided $L_cQ > 2\pi$; if this is
not satisfied then $L_c$ is smaller than the average distance
between solitons and one enters a strong pinning regime.
Following this reasoning\cite{unpub} we obtain an estimate for the critical wavevector
above which one enters the weak pinning regime, 
$Q_c=h\Delta\sqrt{n_p}/2\pi K^2$, where $n_p$ is
the density of pinning centers. This implicitly defines a
critical parallel field scale $B_c$ above which
the depinning drive field $i_c$ will
decrease with $B_{\parallel}$.  Below this, the strings are
strongly pinned by disorder, and  $B_{\parallel}$  has little effect.
This is consistent with our simulation results, as illustrated in Fig. 3.

Our results show that the zero-bias tunneling resonance is not
shifted to higher voltage by an in-plane field, consistent with experiment 
but in contrast
to perturbative theoretical treatments\cite{balents,stern}.  
The latter results suggest that a resonance should be seen at
a voltage corresponding to the
Goldstone collective mode frequency $\hbar \omega(Q) =e V$.
A weak structure {\it is} observed obeying this relation\cite{spielman2}.
While the simulations reported here do not exhibit this behavior, 
it is likely that under different circumstances they would. 
Well above the depinning
threshold, when sliding of the soliton lattice proceeds at
a velocity such that a pinning center crosses a soliton 
once per period of a collective mode, the resulting commensuration
acts as an increased friction that lowers the tunneling
resistance of the system.  A
perturbative analysis\cite{unpub}
supports this, and suggests the width of such a resonance
would be of the order $\Delta V \sim \sqrt{\hbar V\gamma/\Gamma e}$.  For this to be
observable in simulation,
one needs very small values of $\gamma$, requiring impractically
long equilibration times.
Such a
small effective $\gamma$ is
consistent with experiment.

\vskip 5.mm

\begin{figure}
 \vbox to 5.0cm {\vss\hbox to 10cm
 {\hss\
   {\includegraphics{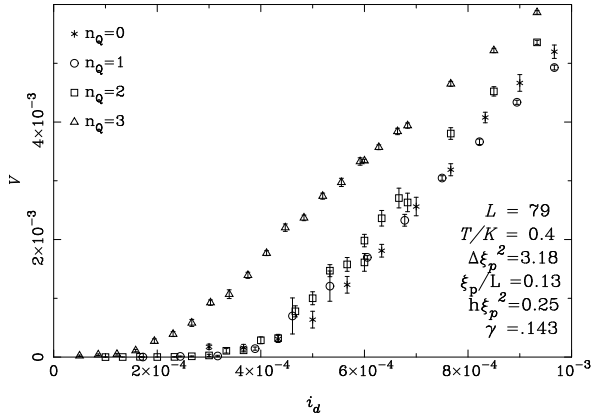}
   }
  \hss}
 }
\vspace{2mm}
\caption{Induced voltage vs. $i_d$ for different parallel magnetic fields.  Units chosen
so that  $\Gamma=K=a_0=1$.}
\label{fig3}
\end{figure}

{\it  Implications and Summary.}  The picture that emerges from
these simulations, and the string glass state
they suggest, has several experimental implications.  These include
the noise peak near the depinning transition and the critical parallel
field discussed above, both of which can in principle be observed.
In addition, one would expect that small systems, for which the
number of vortices generated by disorder is small, would generate
strong sample to sample fluctuations in the tunneling resistance.  
Another signal would be the observation of  a low temperature
scale $T_f \approx \sqrt{Kh}\ell_0$ below which the
isolated excitations freeze out and the tunneling resistance 
drops rapidly\cite{unpub,com3}. $T_f$ is well below experimentally accessible
temperatures for reported sample parameters\cite{spielman1},
but might be observed in samples with large $h$ and $\ell_0$.
Finally, since
vortices carry charge $e/2$ \cite{review} in this system
and are deconfined, it is interesting to speculate that this
fractional charge might be observed in shot noise experiments\cite{com2}.


The authors would like to thank G. Murthy and A.H. MacDonald
for helpful suggestions and discussions.
This work was supported by NSF Grant No.
DMR-0108451.  Computer time was provided by University of
Kentucky through NCSA Grant No. DMR-020024.


\begin{thebibliography}{10}
\bibitem{com1.5} We subtract any $q=0$ component from $A_{\mu}$ to 
guarantee charge neutrality.
\bibitem{com5} A relatively high density of vortices is consistent with 
the quality of samples in Refs. \onlinecite{spielman1,spielman2};
see Ref. \onlinecite{stern}.
\bibitem{com3} This effect is observed in the simulations: for coarse lattices
(small values of $L$), $R$ falls rapidly below a temperature $T_f$.
As $L$ increases with other parameters fixed, $T_f$ becomes small.
\bibitem{com2} This can also occur in clean systems at high enough 
temperature\cite{fertig2,fertig3}.
\end{thebibliography}
\end{document}